\documentclass[journal]{IEEEtran}%
\usepackage{amsmath}
\usepackage{amssymb}
\usepackage{amsfonts}
\usepackage{cite}
\usepackage[]{hyperref}
\usepackage[all]{hypcap}
\usepackage{color}
\usepackage[tight,footnotesize]{subfigure}
\usepackage{comment}
\usepackage{graphicx}%

\title{High Range Resolution Profiling in Missing Data Case: A New Approach}

\author{\IEEEauthorblockN{\emph{Yang Hu, ~Huadong Meng, ~Yimin Liu, ~Xiqin Wang}}

\IEEEauthorblockA{Department of Electronic Engineering,
~Tsinghua University,~Beijing\\
Email: foxsheep@gmail.com}
}

\begin{document}
\maketitle

\begin{abstract}
We have proposed a novel method for Synthetic High Range Resolution (HRR) profiling, under the condition of missing frequency domain samples. This new approach estimates the autocovariance function (ACF) of the signal by valid sample pairs. Autocovariance matrix is formed from ACF estimations. Even with large part of data missing, new approach exhibits robust profiling result. Simulations are presented to show a advantage over other approaches in missing data case. Moreover, a real radar experiment was conducted to validate the new approach.
\end{abstract}

\begin{IEEEkeywords}
Missing Data, Sampled Covariance, HRR
\end{IEEEkeywords}

\section{Introduction}
In profiling radar systems, range resolution is determined by transmitted signal bandwidth. Synthetic bandwidth technique provides high-range-resolution (HRR) capability by transmitting a series of pulses with various carrier frequencies. Within each pulse, the bandwidth is small. The benefit of this method is to achieve large signal bandwidth, while retaining low system complexity. After collecting all the received pulses, the `stretch' algorithm \cite{Einstein1984} can be implemented to derive HRR profiles\cite{wehner1995}. In synthetic HRR profiling, target back scattering property is carefully discussed in \cite{peikang2005}. Theoretical analysis and experimental results had demonstrated that target back scatter can be modeled as distributed point scatterers with individual amplitudes. This approximation of target named `point-scatterer model' is widely recognized in HRR profiling or imaging.

Synthetic bandwidth technique usually suffers from jammers and interferences, due to large bandwidth occupied by transmitted signal. While signal frequencies clashes with environmental electromagnetic interferences, not all of the scattered signal can be correctly acquired by radar receiver. If we let the polluted signal unattended, synthetic HRR profiling will be affected. This frequency domain interference problem in synthetic profiling process can be modeled as `missing data' problem.Various methods have been proposed to interpolate the missing data \cite{Babu2010} \cite{wang2006} \cite{Stoica2009}. These fitting algorithms are based on assumptions of signal models or properties. Then, `stretch' is applied to the processed data. However, if the missing parts are too long, profiling result falls in quality due to inaccuracy interpolation \cite{Babu2010}. For large size of missing, recently developed compressed sensing \cite{Baraniuk2007} \cite{Shah2009} technique is a solution. However, while the missing pattern appears to be block-like, compressed sensing results also deteriorate.

Autocovariance is the second order property of signal. While part of signal is missing, autocovariance function (ACF) of various lags can be estimated from valid data. In this paper, we demonstrate a new approach in missing data case. In which autocovariance matrix is estimated using limited available data. Subspace decomposition method is applied to the estimated matrix to obtain scatterer range information. We present both simulation results and real data from radar systems to validate this new methods. The paper is organized as follows. Section II describes signal model. Section III presents the new method. Simulation and real data result are presented in section IV. Results derived from real radar data is presented in Section V. Section VI concludes the paper.

\section{Signal Model of HRR Profiling}
In synthetic HRR profiling process, radar system transmits a pulse train of $N$ pulses with various carrier frequencies. For the $n$th pulse, carrier frequency is $f_n=f_0+n \Delta f$, where $f_0$ is starting frequency and $\Delta f$ is frequency step. The transmitted pulse for the $n$th frequency is $p_n(t)=A(t)\exp{j2\pi f_n t}$, in which $A(t)$ is the signal envelop. The total bandwidth is $N \Delta f$. Suppose a single reflecting poing with scattering amplitude $\alpha$ is positioning at time delay $\tau$ (corresponding to range at $c \tau /2$). At the receiver, radar received signal will be $r_n(t)=\alpha A(t-\tau)\exp{j2\pi f_n (t-\tau)}$.

In synthetic radar HRR Profiling, point scatterer model describes a target as a series of individual reflecting points \cite{peikang2005} with various amplitudes $[\alpha_1,\alpha_2,\ldots,\alpha_N]$ and time delays $[\tau_1,\tau_2,\ldots,\tau_N]$. These scatterers composes of a linear, time-invariant system. Received signal is the superposition of reflecting wave from all the scatterers, plus noise:
\begin{equation}
r_n(t)=\sum_{k=1}^K \alpha_k A(t-\tau_k)\exp \{j2\pi f_n (t-\tau_k)\}+e(t)
\end{equation}

 After quadrature demodulation and sampling, the complex sample of the $n$th pulse is
\begin{equation} \label{eq:rx}
y_n=\sum_{k=1}^K \alpha_k e^{-j2\pi f_n \tau_k}+e_n
\end{equation}

Obviously, the sampled data is superposition of multiple sinusoids. In synthetic bandwidth system, a series of samples are collected as equally spaced frequency domain samples. With these samples, synthetic HRR profiling is essentially an estimation of three parameters: numbers, time delays and amplitudes. MUSIC \cite{stoica1997} and other subspace methods are used in HRR profiling \cite{Kim2002}. The key step in traditional MUSIC, covariance matrix estimation is operated in full data case.

In synthetic bandwidth signal, due to interference or jamming on the receiver, complex samples may not to be obtainable. Suppose pulse number $P_I=[m_1,m_2,\ldots,m_A]$ are available and $P_N=[m_{A+1},m_{A+1},\ldots,m_N]$ are polluted pulses with interference. We designate 
\begin{IEEEeqnarray*}{C}
I(k)=
  \begin{cases}
    1 & \text{if  } k \in P_I, \\
    0 & \text{if  } k \in P_N.
  \end{cases}
\end{IEEEeqnarray*}
Then $I(k)$ is an indication of the availability of the $k$th pulse. The observations in (\ref{eq:rx}) reduce to:
\begin{equation} \label{eq:rxP}
y_{m_i}=\sum_{k=1}^K \alpha_k e^{-j2\pi f_{m_i} \tau_k}+e_{m_i}
\end{equation}
In the following section, we present a new algorithm named Missing-MUSIC or M-MUSIC, modified for missing data case.

\section{Algorithm Description}

\subsection{Autocovariance Estimation}
We starts the analysis of signal autocovariance function (ACF). For a second-order stationary process, the autocovarainces of different lags do not depend on time shift. In full data case, ACF for a continuously sampled data $X_1,X_2,\ldots,X_N$ with zero mean is estimated by unbiased estimator
\begin{equation}
\hat{c}(h)=\frac{1}{N-h}\sum_{i=1}^{N-h}X_{i+h} X_i^*
\end{equation}
Where $h$ is the lag of ACF and $(\cdot)^*$ is complex transpose. It is stated for lag $h$, we have $N-h$ couples of sampled data to be averaged. For sufficiently long series, the estimation is asymptotic consistent. However, as sampled data is polluted by jammers, the number of available couples will decrease as polluted sample size increases. 

For ACF with lag $h$, the number of useful couples is
\begin{equation}
Q(h)=\sum_{i=1}^{N-h} I(i)I(i+h).
\end{equation}
For each couple of samples that both exist, it will be accounted for ACF estimation. A series of $Q(h)$ for different lags $h$ is calculated. In theory, $Q(h)$ should be in range of $[0,N-h]$.

ACF can be estimated by average all available couples of sampled data in the following manner.
\begin{equation}
\hat{c}(h)=\frac{1}{Q(h)}\sum_{i=1}^{N-h} I(i) I(i+h) X_{i+h} X_i^*.
\end{equation}
Stated simply, we should drop out the couples of polluted samples, average all the `clear' samples left.

\subsection{Covariance Matrix Forming}
In order to proceed subspace identification, a covariance matrix has to be formed. For a second-order stationary process, elements on the same diagonal equal to each other. Thus the covariance matrix should be a Toeplitz matrix, with the value of each diagonal equals to ACF of certain lag.
\begin{equation}
\hat{C}=\left[
\begin{array}{cccc}
\hat{c}(0) & \hat{c}(1) & \ldots & \hat{c}(L-1) \\
\hat{c}(-1) & \hat{c}(0) & \ldots & \hat{c}(L-2) \\
\vdots & \vdots & \ddots & \vdots \\
\hat{c}(1-L) & \hat{c}(2-L) & \ldots & \hat{c}(0)
\end{array}
\right].
\end{equation}
In which $L$ is the size of the covariance matrix. The matrix is essentially both Hermitian and Toeplitz.

The choice of $L$ is important in this method. Too small size will reduce the accuracy and resolution of subspace method \cite{Stoica1989} \cite{Stoica1990}. On the other side, ACF estimation is based on limited samples, missing data condition reduces samples size in advance. Thus the size of $\hat{C}$ should be limited, or ACF estimations of large lags will degenerate covariance matrix property. A rule of thumb is to choose the largest $L$, subject to
\begin{equation}
\forall h<L, \quad Q(h) \ge \frac{N-h}{2}
\end{equation}

\subsection{Number of Scatterers and Range Estimation}
All eigenvalues of estimated covariance matrix $\hat{C}$ is real, since $\hat{C}$ is hermitian. Let $\lambda_1 \ge \lambda_2 \ge \ldots \lambda_L$ denote the eigenvalues of $\hat{C}$ listed in decreasing order. In subspace method such as MUSIC, eigenspaces is divided into signal subspace and noise subspace. Signal subspace is the the span of eigenvectors correspondint to large eigenvalues. In HRR profiling problems, the number of scatterers is usually unknown. However, significant scatterers with strong reflectivity determines the characteristic of radar target. We may determine the number of scatterers by the value of eigenvalues above some `noise level'. Let $T$ be the threshold for significant eigenvalues, the number of scatterers $K$ is the total number of eigenvalues $\alpha_k > T$. $\hat{K}$ is also the dimension of signal subspace, noise subspace dimension is $L-\hat{K}$.

Suppose $\{ s_1,s_2,\ldots,s_{\hat{K}} \}$ are $\hat{K}$ orthogonal eigenvector of the associated $K$ largest eigenvalues of $\hat{C}$. The vectors span the signal space $S$. $\{ g_{\hat{K}+1},g_{\hat{K}+2},\ldots,g_L \}$ are $L-\hat{K}$ eigenvectors spanning the noise subspace $G$. The range or time delay of scatterers is determined by Root-MUSIC \cite{Barabell1983}. If we define the polynomial:
\begin{equation}
g_k (z)=\sum_{l=1}^L g_{lk} z^{-(l-1)}
\end{equation}
Where $g_{lk}$ is the $l$th element of noise-eigenvector $g_k$. Solving the roots of another polynomial:

\begin{equation}
D(z)=\sum_{k=\hat{K}+1}^L g_k(z) s_k^* (1/z^*)
\end{equation}
Each of the polynomial roots has complex form $\hat{z}_i=|\hat{z}_i| e^{j \hat{\omega}_i}$
Obtaining $\hat{K}$ complex roots $\{ Z_1, Z_2, \ldots, Z_{\hat{K}} \}$ that is closest to unit circle on complex plain. The time delay in (\ref{eq:rx}) is
\begin{equation}
\hat{\tau}_i=\frac{\hat{\omega}_i}{2\pi}.
\end{equation}

\subsection{Amplitude Estimation and Profile Forming}
Substituting estimated scatterer number $\hat{K}$ and their time delay $\hat{\tau}_i$ into equation (\ref{eq:rxP}):
\begin{equation}
y_{m_i}=\sum_{k=1}^{\hat{K}} \alpha_k e^{-j2\pi f_{m_i} \hat{\tau}_k}+e_{m_i}
\end{equation}
or in matrix form
\begin{equation}
\mathbf{Y}=\mathbf{\hat{F}}  \alpha + \mathbf{E}
\end{equation}

Each of the estimated scatterers forms a steering vector for observations. The reflective amplitude of $\hat{K}$ scatteres are simply the solution to the least square equations above.
\begin{equation}
\hat{\alpha}=(\mathbf{\hat{F}}^* \mathbf{\hat{F}})^{-1} \mathbf{\hat{F}}^* \mathbf{Y}
\end{equation}

Up to now, all parameters of scatterers are determined. To generate a HRR profile, simply arrange the reflectivity of scatterers according to their time delay or range. Noting the amplitude we estimated is complex. In HRR profiles, amplitude is usually transformed to absolute value and represented in log scale.

\section{Simulation Result}
\begin{figure}[htbp]
\centering
\includegraphics[width=0.48\textwidth]{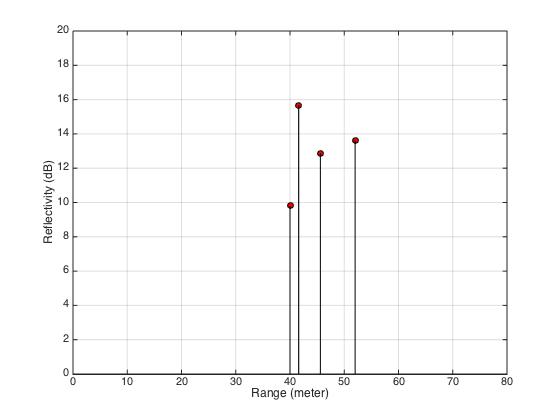}
\caption{Target simulation of four scatterers with different reflectivity. Closest scatterers are separated by 1 meter.}
\label{pic:simutgt}
\end{figure}

To demonstrate the proposed technique, we simulated a synthetic bandwidth system with total bandwidth of $960$MHz, in which $512$ pulses are equally spaced by $\Delta f=1.875$MHz. Theoretical range resolution of this system is about $0.15$ meters. We place four scatterers down range. The locations and amplitudes of scatterers are drawn in figure \ref{pic:simutgt}.

\begin{table}[!t]
\caption{Parameter Table for Missing Data Simulation}
\label{tab:simupara}
\centering
\renewcommand{\arraystretch}{1.1}
\begin{tabular}{c  c}
\hline \hline
Parameter & Value\\
\hline
Radio-frequency & X band\\
Frequency Step Size $(\Delta f)$ & 1.875MHz\\
Total Pulse Number & 512\\
Full Bandwidth & 960MHz \\
Valid Pulse Number & 300\\
SNR at the Receiver & 15dB\\
\hline
\end{tabular}
\end{table}

\subsection{Random Missing Data}

\begin{figure}[htbp]
\centering
\includegraphics[width=0.48\textwidth]{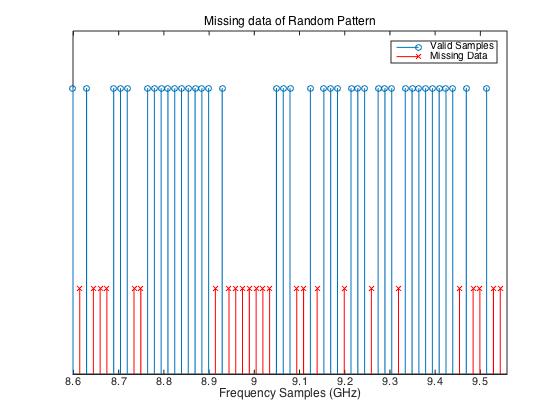}
\caption{Randomly distributed missing data pattern. Blue circles denotes valid sampling frequencies. Red crosses are polluted frequencies. Samples on these frequencies are not used. }
\label{pic:lackrand}
\end{figure}

In this case, interference is randomly distributed on carrier frequencies, as drawn in figure \ref{pic:lackrand}. Simulation parameters are listed in Table \ref{tab:simupara}. Within total bandwidth of 960MHz composed of 512 pulses, 212 pulses are polluted. The unavailable frequencies are equally distributed with in the total bandwidth. We compare the results by compressed sensing and by our new method. Both methods incorporate Akaike information criterion to determine the number of scatterers.

\begin{figure}[htbp]
\centering
\subfigure[]{\includegraphics[width=0.48\textwidth]{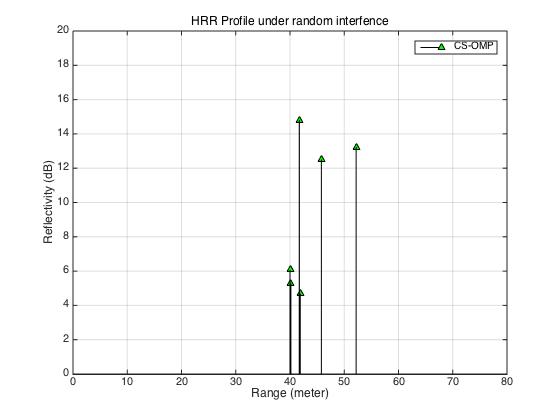} \label{pic:randomp}}
\subfigure[]{\includegraphics[width=0.48\textwidth]{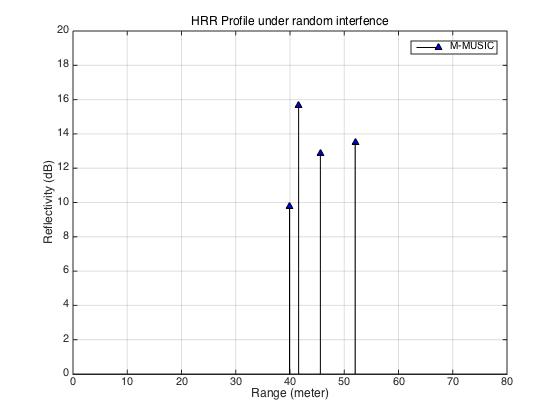} \label{pic:randmusic}}
\caption{(a) HRR profile generated by compressed sensing. (b) HRR profile generated by M-MUSIC. 212 Pulses out of 512 transmitted pulses are polluted and discarded. Polluted frequencies are randomly ditributed over total bandwidth.}
\label{pic:resrand}
\end{figure}

In in figure \ref{pic:resrand}, both result had proved to be capable to profile the scatterers. In both results, the four recovered scatterers appeared at the correct range. Amplitudes also reflect original reflectivity. However, if we look into the detail of the profiling result, new method has displayed some advantages. Compressed sensing result shows more scatterers surrounding the original scatterers. This phenomenon is due to `grid-mismatch' problems in this method \cite{Chi2011}. Meanwhile, new method resolves scatterer range by the root of a polynomial. The range of scatterers are not limited on pre-defined grids.

\subsection{Block Missing Data}
\begin{figure}[htbp]
\centering
\includegraphics[width=0.48\textwidth]{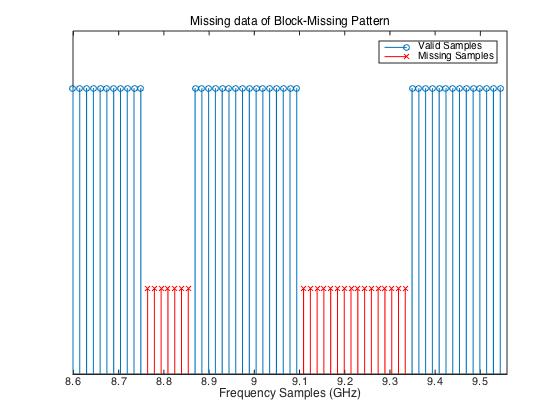}
\caption{Randomly distributed missing data pattern. Blue circles denotes valid sampling frequencies. Red crosses are polluted frequencies. Samples on these frequencies are not used. }
\label{pic:lackblock}
\end{figure}

While the interference over the total bandwidth is block-shaped, as in figure \ref{pic:lackblock}. Simulation parameters are the same as in previous subsection. Only the interference frequency distribution is different. This interference pattern is common in real environment. Signal transmitted by jammers or other radars occupies a continuous part of bandwidth.

\begin{figure}[htbp]
\centering
\subfigure[]{\includegraphics[width=0.48\textwidth]{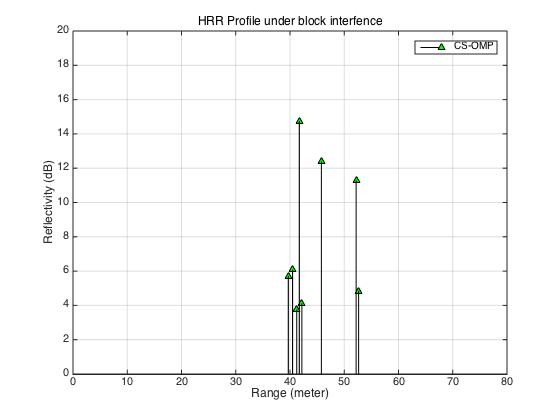} \label{pic:blockomp}}
\subfigure[]{\includegraphics[width=0.48\textwidth]{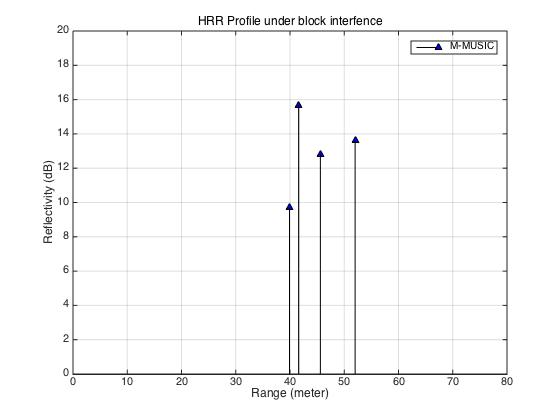} \label{pic:blockmusic}}
\caption{(a) HRR profile generated by compressed sensing. (b) HRR profile generated by M-MUSIC. 212 pulses over two band are interfered.}
\label{pic:resblock}
\end{figure}

The result by compressed sensing has further deteriorated. This is due to the matrix property in this method. The new method displaying stable profiling result exhibit an advantage over traditional method.

\section{Real Radar Data Result}
\begin{figure}[htbp]
\centering
\includegraphics[width=0.48\textwidth]{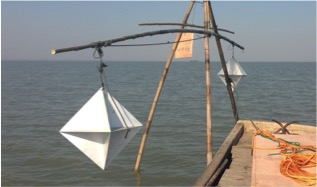}
\caption{Two corner reflectors were placed above water. The range between the reflectors was roughly 4 meters. Two point scatterers are expected at the profile output.}
\label{pic:targetview}
\end{figure}

In order to verify the new method over real system, a synthetic bandwidth radar is placed at the shore of a lake. In this environment, clutter energy and other unwanted interferences are sufficiently low. Experimental data of I/Q channels was collected from the baseband of the radar receiver. The signal parameters are listed in Table \ref{tab:real}. Two corner reflectors separated by about 4 meters are set above the water, as in figure \ref{pic:targetview}. The target is 5km away from radar. In profiling result, we expect to derive two spikes of strong reflection. 200 pulses are deliberately jammed using electronic jammers, in block-like pattern.

\begin{table}[!t]
\caption{Parameter Table for Real System}
\label{tab:real}
\centering
\renewcommand{\arraystretch}{1.1}
\begin{tabular}{c  c}
\hline \hline
Parameter & Value\\
\hline
Radio-frequency & X band\\
Frequency Step Size $(\Delta f)$ & 1.875MHz\\
Intra-pulse Bandwidth & 6MHz \\
Total Pulse Number & 512\\
Full Bandwidth & 960MHz \\
\hline
\end{tabular}
\end{table}

\begin{figure}[htbp]
\centering
\subfigure[]{\includegraphics[width=0.48\textwidth]{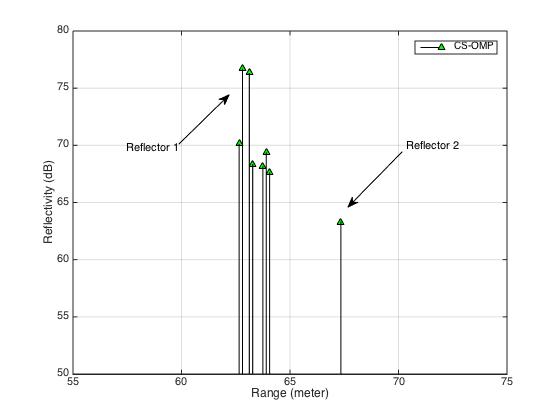} \label{pic:realomp}}
\subfigure[]{\includegraphics[width=0.48\textwidth]{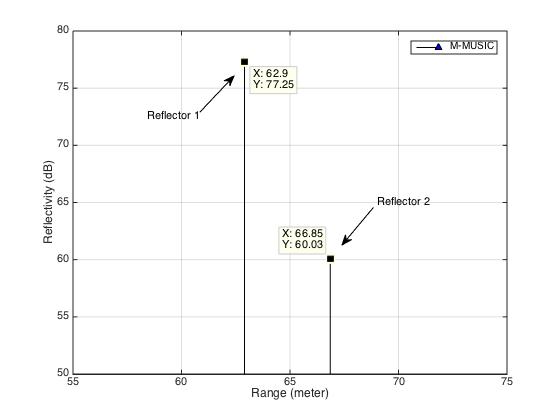} \label{pic:realmusic}}
\caption{Two reflectors are resolved by 3.95 meters. (a) HRR profile by OMP noting multiple reflectors are surrounding the main reflector. This is caused by `Grid-Mismatch'. (b) Result by M-MUSIC. Only two spikes is sufficient to represent the received signal.}
\label{pic:realres}
\end{figure}

We compare compressed sensing and our new method in HRR profile results. In Figure \ref{pic:realomp}, two scatterer points are resolved. However, multiple spikes are required to represent the reflector. This is the phenomenon appeared in simulation data. Result by new method in Figure \ref{pic:realmusic} shows clearly two scatterers. They are separated by 3.95 meters, correctly describing the position of the two reflectors.

\section{Conclusion}
In this paper, we introduced a new HRR profiling algorithm for synthetic bandwidth signal. This new approach is able to correctly resolve scatterer ranges and amplitudes under the condition of missing data. Based on sampled auto-covariance, the signal covariance matrix is formed. Subspace decomposition is applied in order to resolve scatterer ranges. Amplitudes are obtained by least square. Simulations and real data results show the newly method has advantage over compressed sensing method, which is a widely used in missing data case. More application of this method may apply to sinusoid signal decomposition and other radar imaging area.

\bibliographystyle{plain}
\bibliography{HRRP,RadarBasicBooks,ACFEstimation,CovarianceMat,CSTheory,Detection,Mathematics,NonUniform,SpectralAnalysis}

\begin{thebibliography}{10}

\bibitem{Babu2010}
Prabhu Babu and Petre Stoica.
\newblock {Spectral analysis of nonuniformly sampled data – a review}.
\newblock {\em Digital Signal Processing}, 20(2):359--378, March 2010.

\bibitem{Barabell1983}
A~Barabell.
\newblock {Improving the resolution performance of eigenstructure-based
  direction-finding algorithms}.
\newblock {\em Acoustics, Speech, and Signal Processing, IEEE International
  Conference on ICASSP'83.}, pages 8--11, 1983.

\bibitem{Baraniuk2007}
Richard Baraniuk and Philippe Steeghs.
\newblock {Compressive radar imaging}.
\newblock {\em Radar Conference, 2007 IEEE}, pages 128--133, April 2007.

\bibitem{Chi2011}
Yuejie Chi, LL~Louis~L. Scharf, Ali Pezeshki, and A.~Robert Calderbank.
\newblock {Sensitivity to Basis Mismatch in Compressed Sensing}.
\newblock {\em IEEE Transactions on Signal Processing}, 59(5):2182--2195, May
  2011.

\bibitem{Einstein1984}
TH~Einstein.
\newblock {Generation of high resolution radar range profiles and range profile
  auto-correlation functions using stepped-frequency pulse train}.
\newblock Technical report, 1984.

\bibitem{peikang2005}
{Huang Peikang et al.}
\newblock {\em {Radar Target Reflectivity Property}}.
\newblock 雷达技术丛书. Electronic Industry Press, 2005.

\bibitem{Kim2002}
Kyung-tae Kim, Dong-kyu Seo, and Hyo-tae Kim.
\newblock {Efficient Radar Target Recognition Using the MUSIC Algorithm and
  Invariant Features}.
\newblock {\em IEEE Transactions on Antennas and Propagation}, 50(3):325--337,
  March 2002.

\bibitem{Shah2009}
S~Shah, Y~Yu, and A~Petropulu.
\newblock {Step-frequency radar with compressive sampling (SFR-CS)}.
\newblock {\em arXiv preprint arXiv:0910.0886}, (3):1686--1689, 2009.

\bibitem{Stoica1989}
P~Stoica and N~Arye.
\newblock {MUSIC, maximum likelihood, and Cramer-Rao bound}.
\newblock {\em Acoustics, Speech and Signal Processing, IEEE Transactions on},
  17(5), 1989.

\bibitem{stoica1997}
P~Stoica and R~L Moses.
\newblock {\em {Introduction to spectral analysis}}.
\newblock Prentice Hall, 1997.

\bibitem{Stoica1990}
P~Stoica and A~Nehorai.
\newblock {MUSIC, maximum likelihood, and Cramer-Rao bound: further results and
  comparisons}.
\newblock {\em Acoustics, Speech and Signal Processing, IEEE Transactions on},
  3(12), 1990.

\bibitem{Stoica2009}
Petre Stoica, Jian Li, and Hao He.
\newblock {Spectral analysis of nonuniformly sampled data: a new approach
  versus the periodogram}.
\newblock {\em Signal Processing, IEEE Transactions on}, 57(3):843--858, 2009.

\bibitem{wang2006}
Y~Wang, J~Li, and P~Stoica.
\newblock {\em {Spectral Analysis of Signals: The Missing Data Case}}.
\newblock 2006.

\bibitem{wehner1995}
D~R Wehner.
\newblock {\em {High-Resolution Radar}}.
\newblock Radar Library. Artech House, Incorporated, 1995.

\end{thebibliography}
\end{document}